# MuSim, A GRAPHICAL USER INTERFACE FOR MULTIPLE SIMULATION PROGRAMS

Thomas J. Roberts[†], Mary Anne Clare Cummings, Rolland Paul Johnson,
Muons, Inc., Batavia, IL, U.S.A.
David Vincent Neuffer, Fermilab, Batavia, IL, U.S.A.

## *Abstract*

MuSim is a new user-friendly program designed to interface to many different particle simulation codes, regardless of their data formats or geometry descriptions. It presents the user with a compelling graphical user interface that includes a flexible 3-D view of the simulated world plus powerful editing and drag-and-drop capabilities. All aspects of the design can be parameterized so that parameter scans and optimizations are easy. It is simple to create plots and display events in the 3-D viewer (with a slider to vary the transparency of solids), allowing for an effortless comparison of different simulation codes. Simulation codes: G4beamline 3.02 and MCNP 6.1; more are coming. Many accelerator design tools and beam optics codes were written long ago, with primitive user interfaces by today's standards. MuSim is specifically designed to make it easy to interface to such codes, providing a common user experience for all, and permitting the construction and exploration of models with very little overhead. For today's technology-driven students, graphical interfaces meet their expectations far better than text-based tools, and education in accelerator physics is one of our primary goals.

## MuSim [1]

There are dozens of simulation codes in use, and many physicists have complained about the resulting "Tower of Babble"; establishing a common graphical user interface for multiple simulation codes is a major improvement in the field. MuSim can interface to many different beam-optics and particle simulation codes since they necessarily have a common domain, with common concepts, common objects, and common operations. By abstracting these common elements, a single program can indeed interface to many simulation codes with relatively little effort.

Graphical interfaces are used throughout, making it easy to construct the system graphically, display the system with beam tracks, analyze results, and use on-screen controls to vary parameters and observe their effects in (near) real time. Such exploration is essential to give users insight into how systems behave, and is valuable to both experienced and inexperienced scientists, for both teaching and learning, and for tasks such as optimizing a system design using a variety of codes before it is built.

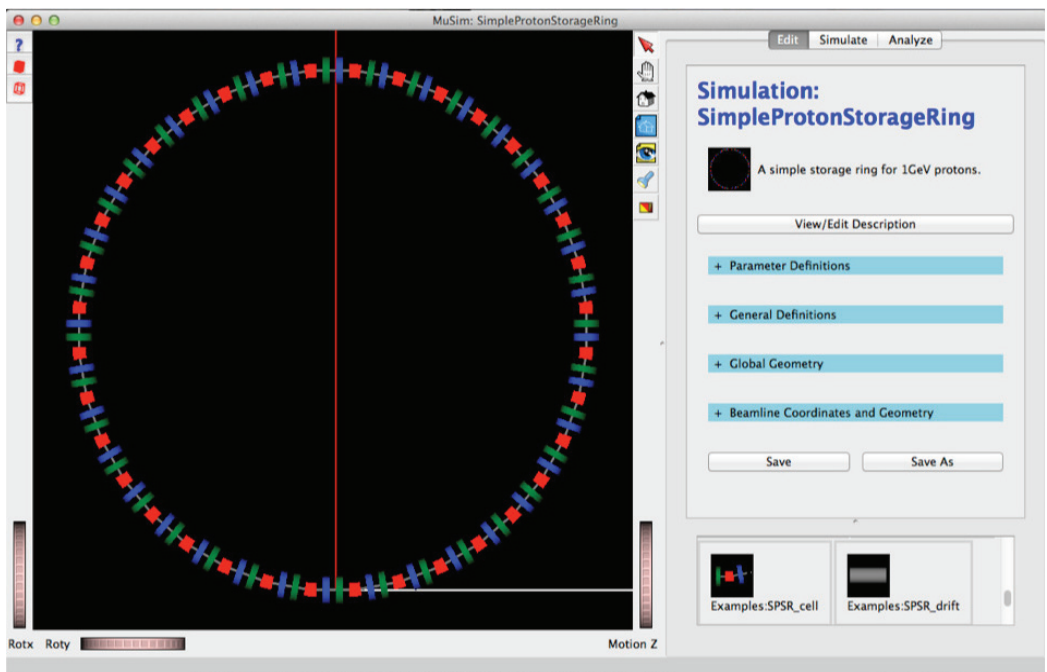

Figure 1: The MuSim main window, showing a 3-D image of the simulated world and an editing panel. At the lower right is the Library; two of its objects are visible – they can be dragged and dropped in the viewer to insert them into the world. This simulation is a simple proton storage ring: dipoles are red, focusing quads are blue, defocusing quads are green; the red and white lines are coordinate axes (the third is out-of-the-page and cannot be seen).

___________
[†] tjrob@muonsinc.com

A primary incentive for developing MuSim has been our interest in accelerator driven subcritical reactors. By combining an accelerator with a nuclear reactor, enormous advantages are gained in terms of both safety and fuel diversity (e.g. it can burn spent nuclear fuel from other reactors *and itself*). Different simulation codes are needed for accelerators and reactors, and they have very different geometry descriptions and data formats; MuSim combines them simply and easily.

MuSim is designed to make simple things simple and complex things possible. While it has a large number of features, they are laid out with usage in mind, making it easy for both novices and experts to navigate. Many users find the online Help to be especially useful, as it automatically follows the user's navigation (click on an object to edit it, or open a dialog box, and the Help window jumps to the page describing that object or dialog box). Figure 1 shows the MuSim main window.

## EDIT, SIMULATE, AND ANALYZE TABS

The *Edit* tab assists the user in creating and editing the simulated world, as well as other aspects of the simulation (e.g. material definitions). The *Simulate* tab permits the user to run simulations using any configured simulation code, including a scan of parameter values and optional sliders for (near) real-time exploration of the effects of changing parameter values; integrated optimization tools are coming. The *Analyze* tab has tools for running analysis programs and generating plots, including graphs of analysis results vs. parameter values. The *Simulate* and *Analyze* tabs can also display event tracks in the 3-D viewer; there is a slider to control the transparency of all solids, so tracks can easily be seen inside or behind them.

## UNITS

As different simulation codes use different systems of units, MuSim includes a units-aware expression evaluator. Compatible units can be intermixed in any valid combination (invalid usage is flagged as an error):

1 cm+1 in=3.54 cm; 1 ft/1 cm=30.48; 17*1 ns=17 ns

The edit field for a parameter with units (e.g. a length) displays a pull-down list of all compatible units. When necessary, any expression can be converted to a specific (compatible) unit.

Units can be intermixed freely, and the geometry of elements placed into the world can use any length unit; EM fields can use any appropriate unit, etc. The module for each simulation code knows the system of units it uses and performs conversions automatically.

## LIBRARIES

Users can construct customized objects appropriate for their simulations. These can be collected into one or more libraries, from which the objects can be inserted into simulations (or into other objects) via simple drag-and-drop, or via manual insertion. A library can be accessed via a URL, permitting efficient collaboration among groups distributed around the world.

## EXAMPLES

A major aspect of MuSim is its large collection of examples. These serve as both a demonstration of how to use various MuSim features, as well as being starting points for users to build their own simulations. They are quite varied, ranging from a simple beam going through an absorber, to a simple proton storage ring (shown in Fig. 1 above), to a complex accelerator driven subcritical reactor. Figures 2 – 7 show a few of them.

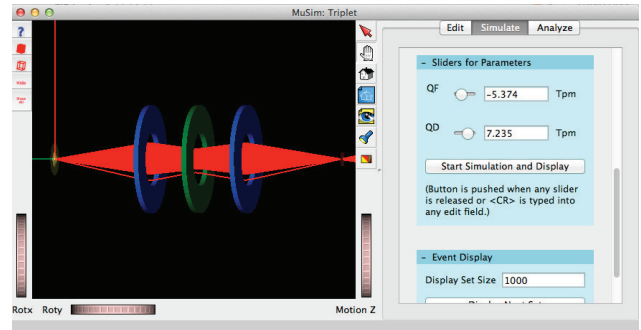

Figure 2: A quad triplet with sliders for field values. Changing a value causes the simulation to be re-run and the tracks (red) to be re-displayed (typically 1-2 seconds).

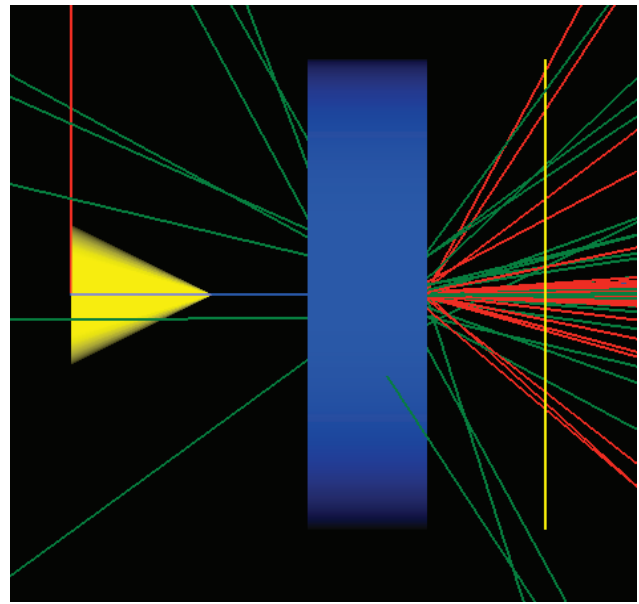

Figure 3: A 200 MeV electron beam hits a cylinder of $H_2O$ (blue), followed by a detector (yellow disk seen edge-on). The beam emanates from the yellow cone at the left. Tracks: electrons are red and gammas are green.

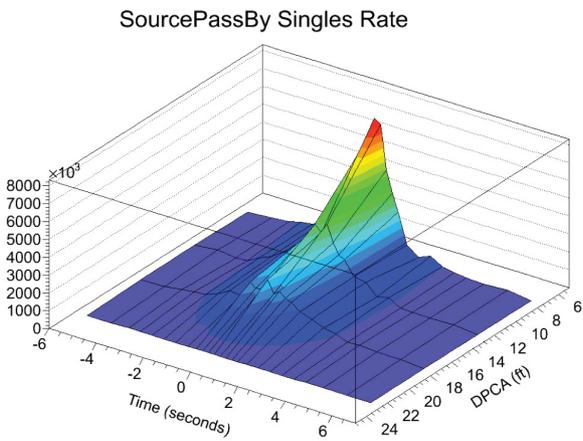

Figure 4: Plot of a roadside detector singles rate as a 1 Ci $^{60}$Co source passes by at 10 MPH, for various distances to the point of closest approach (DPCA). This example can be easily adapted to add various types of shielding.

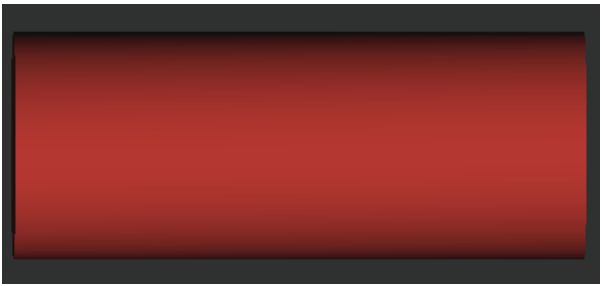

Figure 5: A $^{238}$Pu radioisotope Thermal Generator. The power output is easily plotted. Radioactive decay is one of MuSim's many sources.

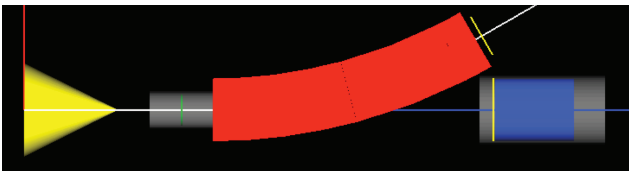

Figure 6: The radiator portion of an energy-recovery linac for radioisotope production. The electron beam emanates from the yellow cone, hitting the gamma production target (green) inside a beam pipe (gray); the charged particles are bent by the magnet (red), while the gammas exit through a hole in its side along the blue line to the production target (blue).

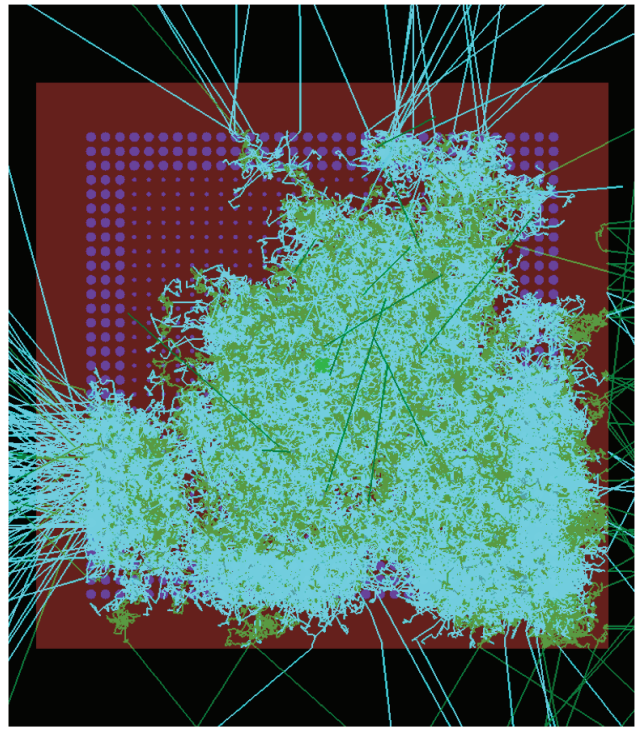

Figure 7: A simple accelerator-driven subcritical reactor. This shows 20,000 tracks (out of 585,000) from a single 1 GeV proton. The beam is from G4beamline [2] and the reactor simulation is MCNP 6.1 [3]. Neutrons are green, photons are cyan, graphite is brown, and the molten salt fuel is blue. This is an end view with the beam coming out of the page. Solids are 50% transparent to show the tracks.

## REFERENCES

[1] MuSim, http://musim.muonsinc.com
[2] G4beamline, http://g4beamline.muonsinc.com
[3] MCNP, http://mcnp.lanl.gov